\documentclass{ws-procs9x6}

\usepackage{psfrag}
\usepackage{graphicx}
\usepackage{dcolumn}

\newcommand{\dto}{D$_{2}$O}
\newcommand{\hto}{H$_{2}$O}

\newcommand{\nue}{$\nu_{e}$}
\newcommand{\numu}{$\nu_{\mu}$}
\newcommand{\nutau}{$\nu_{\tau}$}
\newcommand{\nux}{$\nu_{x}$}

\newcommand{\teff}{$T_{\rm eff}$}
\newcommand{\costs}{$\cos\theta_{\odot}$}

\newcommand{\snoncfluxunc}{6.42^{+1.57}_{-1.57}\mbox{(stat.)}^{+0.55}_{-0.58}~\mbox{(syst.)}} 
\newcommand{\snomutauflux}{3.41^{+0.45}_{-0.45}\mbox{(stat.)}^{+0.48}_{-0.45}~\mbox{(syst.)}} 
\newcommand{\snoeflux}{1.76^{+0.05}_{-0.05}\mbox{(stat.)}^{+0.09}_{-0.09}~\mbox{(syst.)}}

\newcommand{\nsigmassno}{5.3}

\newcommand{\saltccfluxunc}{1.59^{+0.08}_{-0.07}\mbox{(stat)}^{+0.06}_{-0.08}\mbox{(syst)}} 
\newcommand{\saltesfluxunc}{2.21^{+0.31}_{-0.26}\mbox{(stat)}~\pm{0.10}~\mbox{(syst)}} 
\newcommand{\saltncfluxunc}{5.21\pm 0.27~\mbox{(stat)}~\pm0.38~\mbox{(syst)}}

\newcommand{\saltnccfit}{1339.6^{+63.8}_{-61.5}} 
\newcommand{\saltnesfit}{170.3^{+23.9}_{-20.1}} 
\newcommand{\saltnncfit}{1344.2^{+69.8}_{-69.0}}  
\newcommand{\saltnncbefit}{84.5^{+34.5}_{-33.6}}
\newcommand{\saltnsigmamax}{5.4}

\begin{document}

\title{Solving the Solar Neutrino Problem 2~\lowercase{km} Underground --- the Sudbury Neutrino Observatory}

\author{A.W.P. Poon\footnote{\uppercase{O}n behalf of the \uppercase{SNO  C}ollaboration}}

\address{Institute for Nuclear and Particle Astrophysics, \\ Lawrence Berkeley National Laboratory\\
1 Cyclotron Road, Berkeley, CA 94720, USA \\
E-mail: awpoon@lbl.gov}

\maketitle

\abstracts{
The Sudbury Neutrino Observatory (SNO) is capable of measuring simultaneously the flux of electron-type neutrinos and the total flux of all active flavours of neutrinos originating from the Sun.  A model-independent test of neutrino flavour transformation was performed by comparing these two measurements.  Assuming an undistorted neutrino energy spectrum, this transformation has been definitively demonstrated in the pure \dto\ phase of the SNO experiment.  In the second phase with dissolved NaCl in the \dto, the total active solar neutrino flux was measured without any assumption on the energy dependence of flavour transformation.  In this talk, results from these measurements, their physics implications and the current status of the SNO experiment are presented.
}

\section{Introduction}
\label{sec:intro}

For over three decades, solar neutrino experiments\cite{homestake,kamioka,sage,gallex,gno,superk}
have been observing fewer neutrinos than what are predicted by the
detailed models (e.g. \cite{bpb}) of the Sun.  This deficit is known as the Solar Neutrino Problem.   These experiments probe different parts of the solar neutrino energy spectrum, and show an energy dependence in the measured solar neutrino flux.  These observations can be explained if the solar models are incomplete or neutrinos undergo flavour transformation while in transit to the Earth. 

The Sudbury Neutrino Observatory\cite{sno}, using 1000 tonnes of
99.92\% isotopically pure \dto\ as the neutrino target, was constructed to resolve this puzzle.  The \dto\ target is contained in a 12-m diameter acrylic sphere, which is surrounded by 7000 ~tonnes of ultra-pure \hto.  This volume of
\hto\ shields the \dto\ from high-energy $\gamma$ rays and neutrons
originating from the cavity wall.  A 17.8-m diameter stainless steel
structure supports 9456 20-cm diameter inward-facing photomultiplier tubes
(PMTs).  A non-imaging light concentrator is mounted on each PMT, extending the total photocathode coverage to 55\%.  An additional 91 PMTs are
mounted facing outward on the support structure to serve as a cosmic-ray 
veto.  

The Sudbury Neutrino Observatory detects solar neutrinos by
\[\begin{array}{lcll}
    \nu_{e}+d & \rightarrow & p+p+e^{-} & \hspace{0.5in} \mbox{(CC)}\\ 
    \nu_{x}+d & \rightarrow & p+n+\nu_{x} & \hspace{0.5in} \mbox{(NC)} \\
    \nu_{x}+e^{-} & \rightarrow &  \nu_{x}+e^{-} & \hspace{0.5in} \mbox{(ES).} \\
\end{array}\]
The charged-current (CC) reaction on the deuteron is sensitive exclusively to \nue, 
and the neutral-current (NC) reaction has equal sensitivity to all 
active neutrino flavours (\nux ; $x=e,\mu,\tau$).  Elastic scattering 
(ES) on electron is also sensitive to all active flavours, but with 
reduced sensitivity to \numu\ and \nutau.  By counting the free neutron in the final state of the NC reaction, the total active $^8$B neutrino flux can be inferred for neutrinos with energy above the 2.2-MeV kinematic threshold.   

In the first phase of the SNO experiment, the free neutron from the NC interaction was detected by observing the 6.25-MeV $\gamma$ ray following its capture on a deuteron.  To enhance the neutron detection efficiency, 2 tonnes of salt was added in the second phase of the experiment.  The free neutron was captured by $^{35}$Cl and a $\gamma$-ray cascade with a total energy of 8.6~MeV is emitted.  In the final phase of the experiment, $^3$He proportional counters, the ``Neutral-Current Detectors" (NCD)\cite{browne}, are deployed on a 1-m grid in the \dto.  Free neutrons are captured by $^3$He.  In this talk, solar neutrino results from the first two phases of the SNO experiment\cite{snocc,snonc,snodn,snosaltnc} are presented, and their physics implications are discussed.  

\section{Previous Results from the Pure \dto\ Phase}
\label{sec:d2o}

Data from the first phase of the experiment, in which a pure \dto\ target was used, correspond to a livetime of 306.4~days.  Over 450~million triggers were recorded during this period. 

To test the null hypothesis of no neutrino flavour transformation, the extended maximum likelihood method was used to extract the CC, ES and neutron (i.e. NC+background) contributions in the data set.  Backgrounds were constrained to the measured values.  Data distributions in \teff,  the volume-weighted radial 
variable $(R/R_{AV})^{3}\equiv\rho$ and \costs\ were simultaneously fitted to the probability density functions (PDFs) 
generated from simulations.  $R_{AV}=600$~cm is the radius of the acrylic vessel, and $\theta_\odot$ is the angle between the reconstructed direction of the event and the instantaneous direction from the Sun to the Earth.   Assuming an undistorted $^{8}$B energy spectrum\cite{ortiz},  the fluxes of the \nue\ ($\phi_e$) and non-\nue\ ($\phi_{\mu\tau}$) components were found to be (in units of 10$^6$~cm$^{-2}$~s$^{-1}$):
\begin{eqnarray*}
\phi_{e} & = & \snoeflux \\
\phi_{\mu\tau} & = & \snomutauflux 
\end{eqnarray*}
Combining the statistical and systematic uncertainties in quadrature, $\phi_{\mu\tau}$ is \nsigmassno$\sigma$  above zero, and provides strong evidence for flavour transformation.  Removing the energy constraint, the signal decomposition was repeated using only the $\cos \theta_{\odot}$ and $\rho$.  The total flux of active ${}^{8}$B neutrinos in the unconstrained fit was found to be (in units of 10$^6$~cm$^{-2}$~s$^{-1}$)
\begin{displaymath}
\phi_{\mbox{\tiny NC}}=  \snoncfluxunc 
\end{displaymath}

\section{The Salt Phase}
\label{sec:salt}

Neutrino flavour transformation can be an energy dependent process (e.g. the MSW mechanism\cite{msw}).  The energy constrained analysis in pure \dto\ made the explicit assumption of an undistorted $^8$B neutrino spectrum.  The unconstrained results suffered from the strong anti-correlation between the CC and NC channels.  The addition of 2 tonnes of NaCl to the D$_2$O target was aimed at improving the separation between CC and NC without invoking the energy constraint.  Neutron capture on $^{35}$Cl typically produces multiple $\gamma$ rays while the CC and ES reactions produce single electrons. The greater isotropy of the Cherenkov light from neutron capture events relative to CC and ES events allows good statistical separation of the event types.  

Data from the salt phase that are presented here were recorded between July 26, 2001 and October 10, 2002, totaling 254.2 live days.  The number of triggers recorded was over 435~million.   To reduce the effects of backgrounds in the analysis, an effective electron kinetic energy threshold $T_{\rm eff}$ $\geq$ 5.5 MeV and a fiducial volume
with radius $R_{\rm{fit}} \leq 550$ cm were chosen.  

Detector diagnostic triggers and instrumental background events were first removed in the offline data processing.  The latter reduction was accomplished by a set of algorithms that  remove events that do not have the characteristics of Cherenkov light emission.  For the events that passed the first offline reduction, the calibrated times and positions of the hit photomultiplier tubes (PMTs) were used to reconstruct the vertex position and the direction of the particle.  The energy estimator then assigned an effective kinetic energy \teff\ to each event based on these reconstructed parameters and the number of hit PMTs.  The energy estimator used calibration results such as the time variation of light attenuation at various wavelengths as inputs, and was normalized to the detector response to a $^{16}$N source\cite{nsix}.  ${}^{16}$N data taken throughout the running period verified the gain drift ($\sim$2\% year$^{-1}$) predicted by simulations based on the optical measurements.  The energy scale uncertainty is $1.1\%$.   After all the data reduction cuts and the selections of energy and fiducial volume, the data set was reduced to 3055 candidate events.

\begin{figure}
\begin{center}
\centerline{\epsfxsize=3.25in\epsfbox{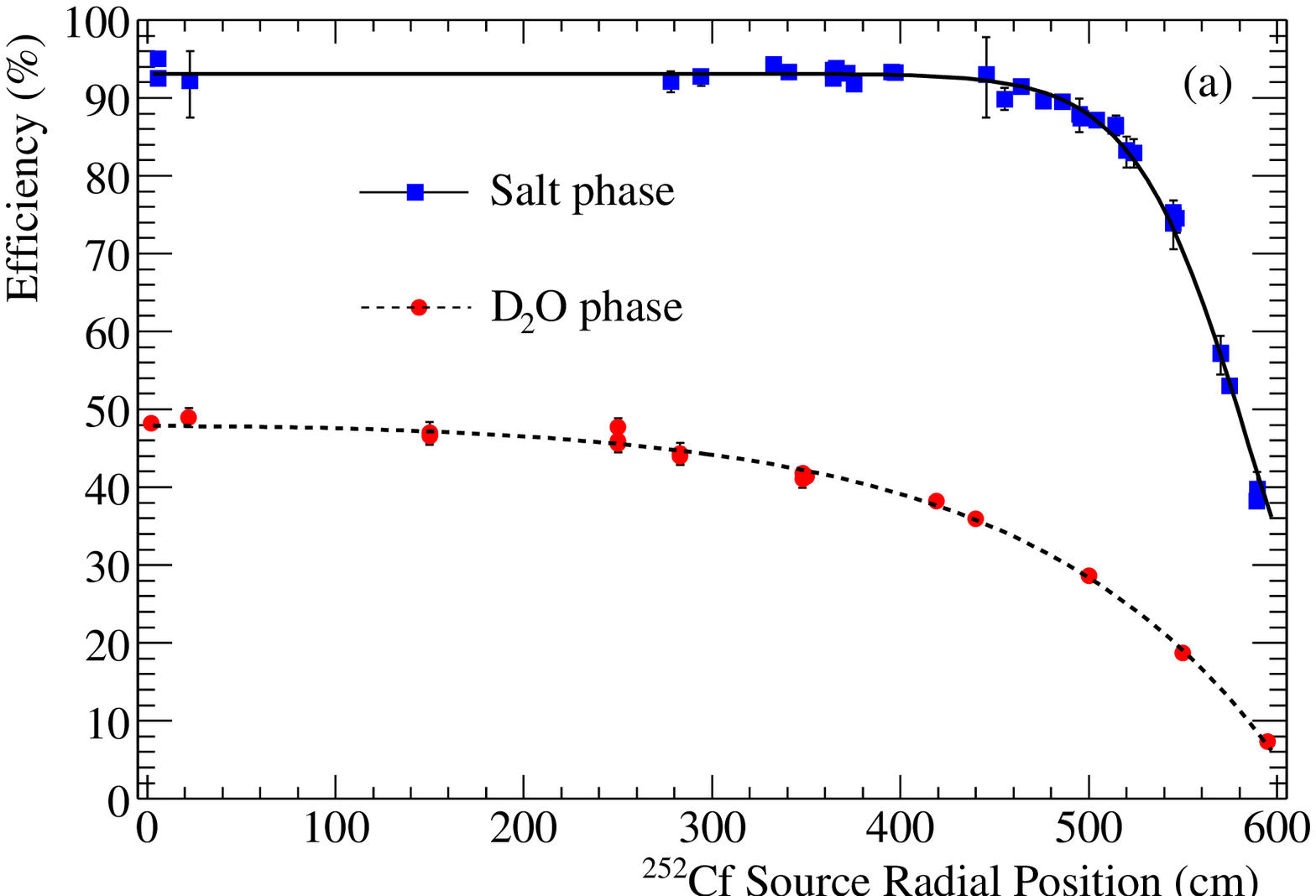}}   
\centerline{\epsfxsize=3.25in\epsfbox{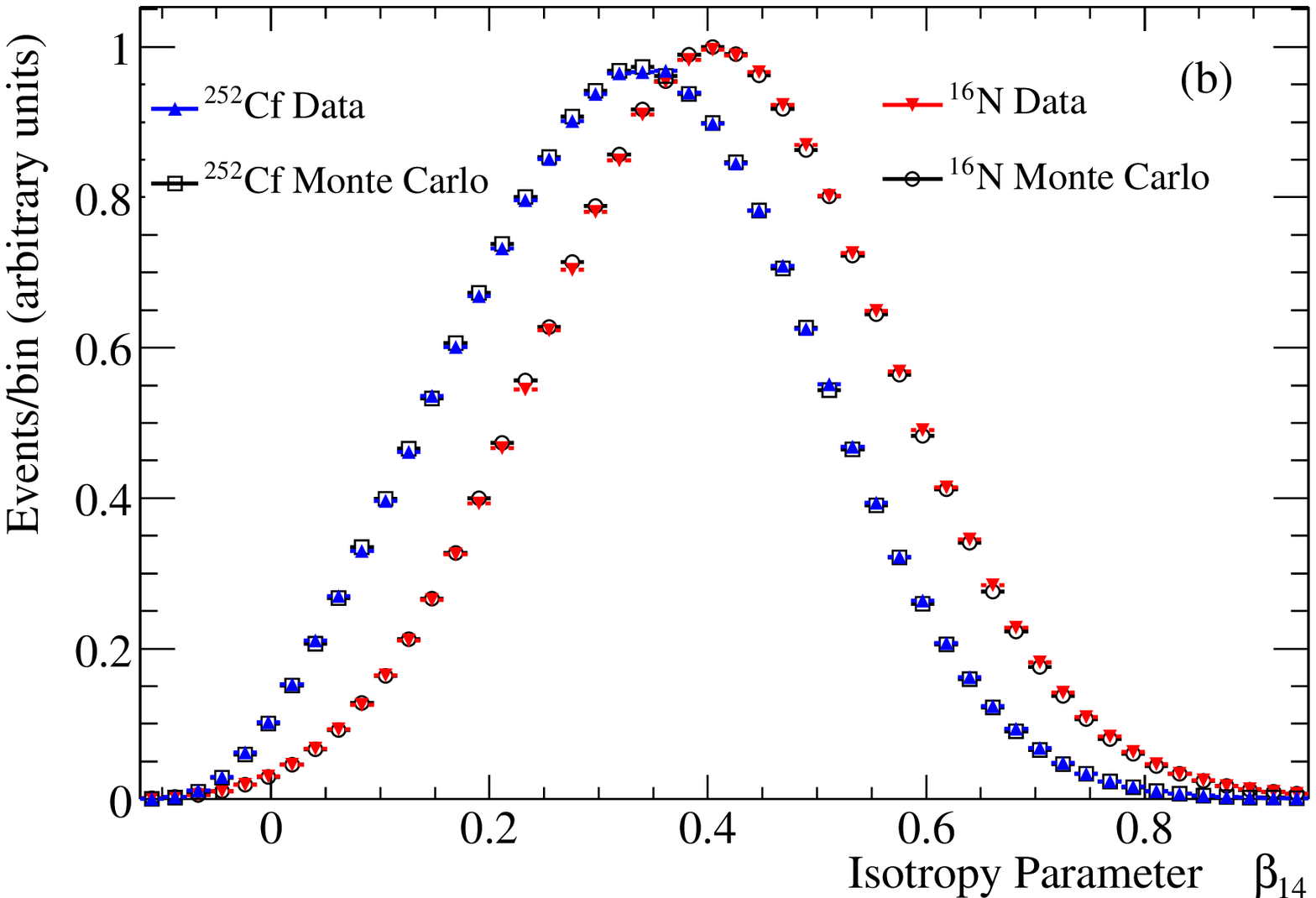}}   
\caption{\label{fig:salt1}(a) Neutron capture efficiency versus source radial position for the pure D$_2$O phase (capture on D) and salt phase (capture on Cl or D) deduced from a ${}^{252}$Cf source, with fits to an analytic function (salt) and to a neutron diffusion model (D$_2$O). (b) Event isotropy from data and Monte Carlo calculations of a ${}^{252}$Cf source and an ${}^{16}$N $\gamma$-ray source.}
\end{center}
\vspace{-4ex}
\end{figure}

The neutron detection efficiency was calibrated with a $^{252}$Cf fission source.  The detection 
efficiency for NC reactions in the heavy water was $0.399 \pm 0.010~\mbox{(calibration)}\pm0.009~\mbox{(fiducial volume)}$ for the chosen energy threshold and fiducial volume, an increase of approximately a factor
of three from the pure \dto\ phase.  This is shown in Fig.~\ref{fig:salt1}(a).

 Event isotropy was characterized by $\beta_{l}$, the mean 
 of the Legendre polynomial $P_{l}$ of the cosine of the angle between PMT hits $\theta_{ij}$ :
 \begin{equation*}
 \beta_l = \frac{2}{N(N-1)}\sum_{i=1}^{N-1} \sum_{j=i+1}^N
  P_l({\rm cos}\ \theta_{ij})
 \end{equation*}
 The combination $\beta_{1}+4\beta_{4} \equiv \beta_{14}$ was selected as the isotropy measure to separate NC and CC events.  Systematic uncertainty on $\beta_{14}$ distributions for simulated 
signal events was evaluated by comparing ${}^{16}$N calibration data to simulations for events
throughout the fiducial volume and running period. The uncertainty on the mean value of $\beta_{14}$ is $0.87\%$.  Comparisons of $\beta_{14}$ distributions from ${}^{16}$N events and neutron events from ${}^{252}$Cf to 
Monte Carlo calculations are shown in Fig.~\ref{fig:salt1}(b).  The Monte Carlo calculations of $\beta_{14}$ have been verified with a 19.8-MeV $\gamma$-ray source \cite{poon}, high-energy electron events (dominated by CC and ES interactions) from the pure \dto\ phase, neutron events following muons, and low-energy calibration sources.

\begin{table}[pt]
\tbl{Background events. The internal neutron and $\gamma$-ray backgrounds are constrained in the analysis. The external-source neutrons are reported from the fit.\vspace*{1pt}}
{\footnotesize
\begin{tabular}{ll} \hline
Source                                  & Events                        \\ \hline
Deuteron photodisintegration              &  $73.1^{+24.0}_{-23.5}$               \\
${}^{2}$H($\alpha,\alpha$)$pn$            &  $2.8 \pm 0.7$                  \\
${}^{17,18}$O($\alpha$,$n$)                  &  $1.4 \pm 0.9$                       \\
Fission, atmospheric $\nu$ (NC +                    &                                    \\
\hspace{2em}sub-Cherenkov threshold CC)         &  $ 23.0 \pm 7.2$                   \\
Terrestrial and reactor $\bar{\nu}$'s   &  $2.3 \pm 0.8$                \\
Neutrons from rock                      &  $\leq 1$                       \\ 
${}^{24}$Na activation                  &  $8.4 \pm 2.3$                   \\
$n$ from CNO $\nu$'s                             &  $0.3 \pm 0.3$                   \\ \hline
Total internal neutron background       &  $111.3^{+25.3}_{-24.9}$               \\ \hline
Internal $\gamma$ (fission, atmospheric $\nu$) & $5.2 \pm 1.3$ \\  
${}^{16}$N decays                       & $< 2.5$ (68\% CL) \\ \hline
External-source neutrons (from fit)  &  $84.5^{+34.5}_{-33.6}$ \\ \hline
Cherenkov events from $\beta-\gamma$ decays   & $<14.7$  (68\% CL) \\ 
``AV events''                                          & $<5.4$ (68\% CL) \\  \hline
\end{tabular}\label{tab:le}}
\vspace*{-13pt}
\end{table}

Radioactive decays of the daughters in the natural $^{232}$Th and $^{238}$U chains were the dominant backgrounds in the neutrino signal window.  The levels of  $^{232}$Th and $^{238}$U daughters in the detector were determined by radiochemical assays of the detector target\cite{ander} and by analyzing the Cherenkov signals from their decays in the neutrino data set.  The average rate of background neutron production due to photodisintegration of the deuteron from radioactivity in the \dto\ is  $0.72 ^{+0.24}_{-0.23}$ neutrons per day.   $^{24}$Na  was produced in the water circulation system by neutrons emitted from the rock wall in the underground laboratory.  It could also be produced {\em in-situ} by calibration sources (e.g. the $^{252}$Cf source).  In order to understand its production and distribution, a controlled neutron activation was carried out.  The amplitude of this background was found to be small.  

Neutrons and $\gamma$ rays produced at the acrylic vessel and in the \hto\ can propagate into the fiducial volume.  Radon progeny deposited on the surfaces of the vessel during construction can initiate ($\alpha$,$n$) reactions on  $^{13}$C, $^{17}$O, and $^{18}$O, and external $\gamma$ rays can photodisintegrate the deuteron giving rise to a free neutron.  The radial distribution of these external-source neutrons was included in the analysis to extract this component.  A class of background events were reconstructed near the acrylic vessel and were characterized by a nearly isotropic light distribution (``AV events").  A limit of 5.4 events (68\% CL) was set for this background.  Table~\ref{tab:le} is a summary of the backgrounds.

\begin{figure}[ht]
\begin{center}
\psfrag{COSSUN}{~$\cos \theta_{\odot}$}
\centerline{\epsfxsize=2.75in\epsfbox{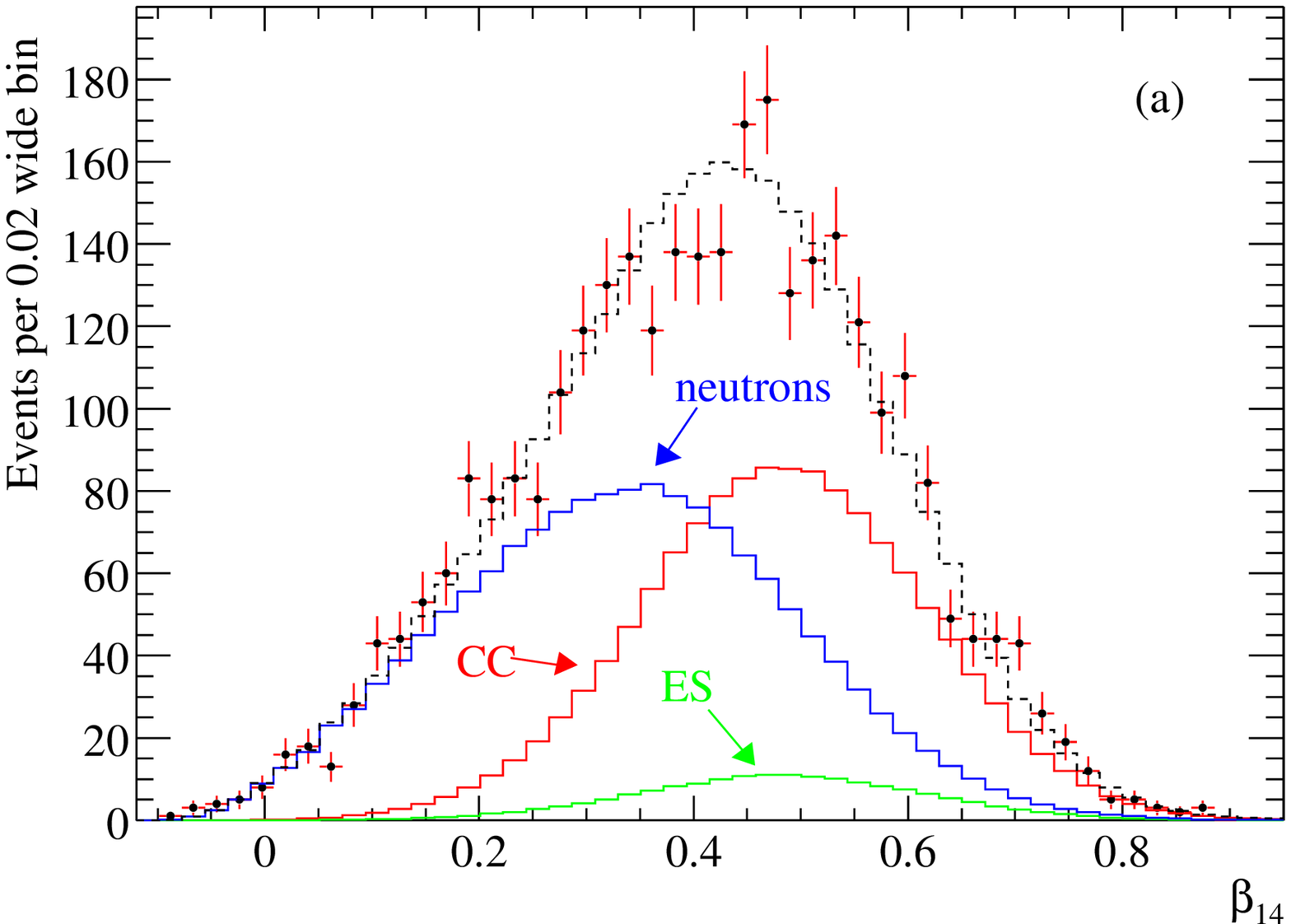}}   
\centerline{\epsfxsize=2.75in\epsfbox{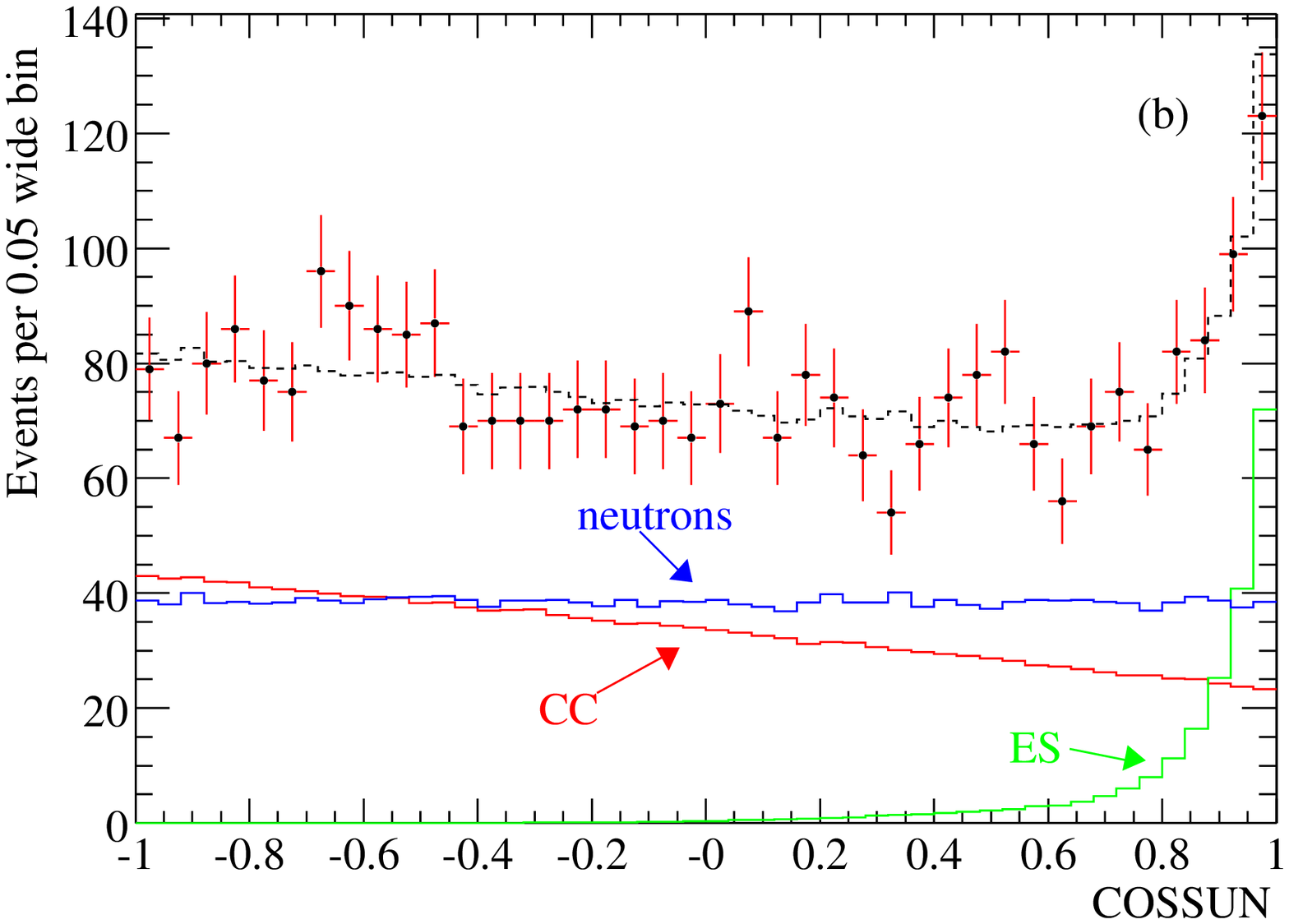}} \vspace{0.2in} \hspace{-0.26in}
\centerline{\epsfxsize=2.5in\epsfbox{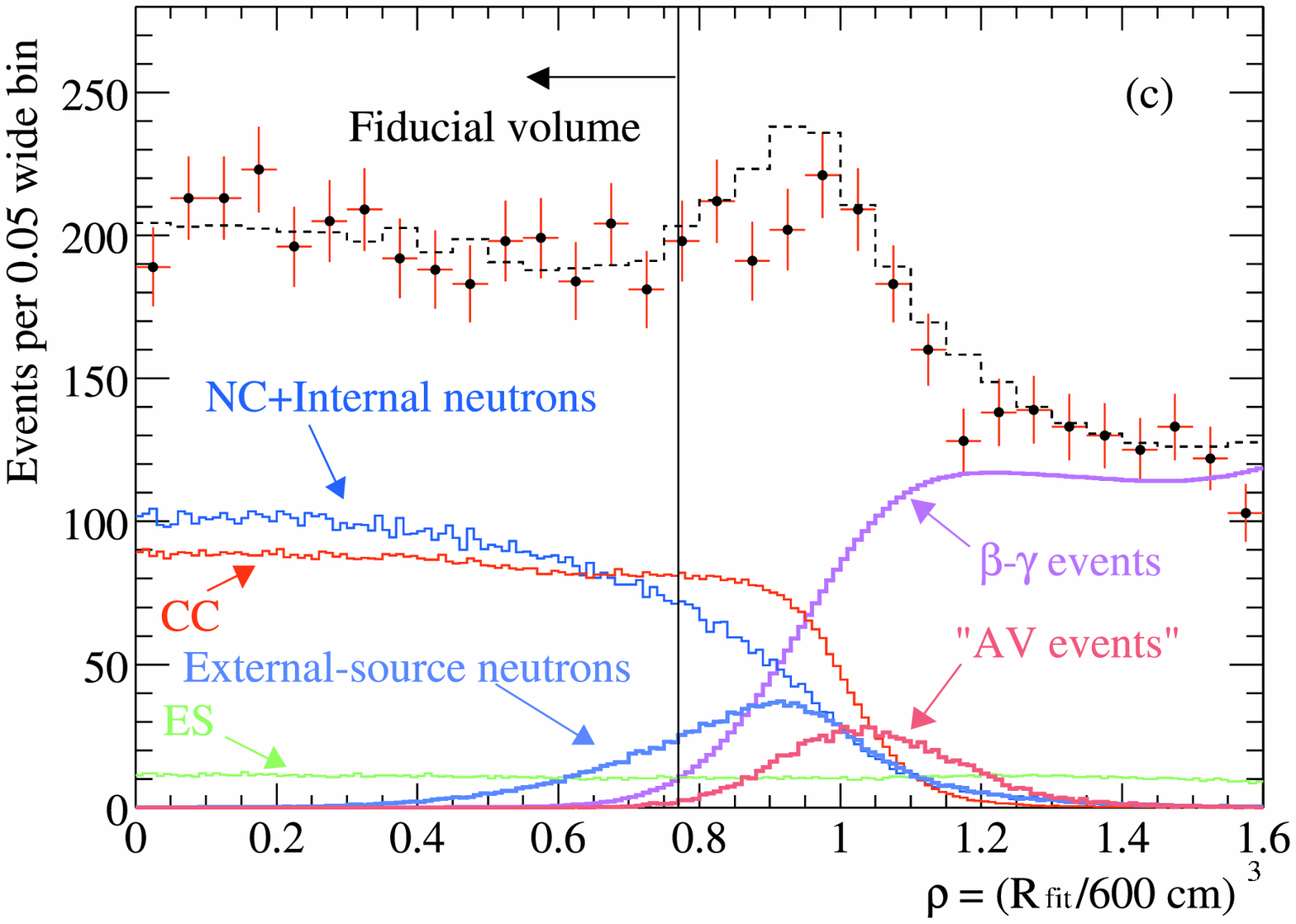}}   
\caption{\label{fig:salt_data}Distribution of (a) $\beta_{14}$, (b) $\cos \theta_{\odot}$ and (c) volume-weighted radius  ($\rho$) distribution.  Also shown are the Monte Carlo predictions for CC, ES, NC + internal and external-source neutron events, all scaled to the fit results.  The dashed lines represent the summed components.  All fits were done with $T_{\rm eff}$$\geq$5.5 MeV and R$_{\rm fit}$$\leq$ 550 cm. }
\end{center}
\end{figure}

A blind analysis procedure was used to minimize the possibility of introducing biases. The data set used during the development of the analysis procedures and the definition of parameters excluded an unknown fraction ($<$30\%) of the final data set, included an unknown admixture of muon-following neutron events, and included an unknown NC cross-section scaling factor.  After fixing all analysis procedures and parameters, the blindness constraints were removed. The analysis was then performed on the ``open" data set, statistically separating events into CC, NC, ES, and external-source neutrons using an extended maximum likelihood analysis based on the 
distributions of isotropy, \costs, and $\rho$.  The fits are shown, with statistical uncertainties only, in Fig.~\ref{fig:salt_data}.  This analysis differs from the analysis reported in Sec.~\ref{sec:d2o} that no assumption on the energy dependence of the neutrino flavour transformation was made.    

The extended maximum likelihood analysis  yielded $\saltnccfit$ CC, $\saltnesfit$ ES,
$\saltnncfit$ NC, and $\saltnncbefit$ external-source neutron events.  The systematic uncertainties on derived fluxes are shown in Table~\ref{tab:errors}.  The fitted numbers of events give the equivalent $^8$B
fluxes~(in units of $10^6~{\rm cm}^{-2} {\rm s}^{-1}$): 
\begin{eqnarray*}
\phi^{\text{SNO}}_{\text{CC}} & = & \saltccfluxunc \\
\phi^{\text{SNO}}_{\text{ES}} & = & \saltesfluxunc \\
\phi^{\text{SNO}}_{\text{NC}} & = & \saltncfluxunc~\mbox{.}
\end{eqnarray*} 
An analysis with the constraint of an undistorted ${}^{8}$B energy spectrum yielded results which are consistent with previous results in Sec~\ref{sec:d2o}.

\begin{table}
\tbl{Systematic uncertainties on fluxes for the spectral shape unconstrained analysis of the salt data set. $\dagger$ denotes CC vs NC anti-correlation. }
{\footnotesize
\begin{tabular}{llllll}\hline
  Source       & NC uncert. & CC uncert. & ES uncert. \\ 
         & (\%) & (\%) & (\%) \\  \hline 
    Energy scale   & -3.7,+3.6 & -1.0,+1.1 & $\pm1.8$ \\ 
Energy resolution   & $\pm1.2$ & $\pm0.1$ & $\pm0.3$ \\ 
Energy non-linearity   & $\pm0.0$ & -0.0,+0.1 & $\pm0.0$ \\ 
 Radial accuracy   & -3.0,+3.5 & -2.6,+2.5 & -2.6,+2.9 \\ 
Vertex resolution   & $\pm0.2$ & $\pm0.0$ & $\pm0.2$ \\ 
Angular resolution   & $\pm0.2$ & $\pm0.2$ & $\pm2.4$ \\ 
   Isotropy mean $\dagger$  & -3.4,+3.1 & -3.4,+2.6 & -0.9,+1.1 \\ 
Isotropy resolution   & $\pm0.6$ & $\pm0.4$ & $\pm0.2$ \\ 
Radial energy bias   & -2.4,+1.9 & $\pm0.7$ & -1.3,+1.2 \\ 
Vertex Z accuracy $\dagger$  & -0.2,+0.3 & $\pm0.1$ & $\pm0.1$ \\ 
Internal background neutrons   & -1.9,+1.8 & $\pm0.0$ & $\pm0.0$ \\ 
Internal background $\gamma$'s   & $\pm0.1$ & $\pm0.1$ & $\pm0.0$ \\ 
 Neutron capture   & -2.5,+2.7 & $\pm0.0$ & $\pm0.0$ \\ 
Cherenkov backgrounds   & -1.1,+0.0 & -1.1,+0.0 & $\pm0.0$ \\ 
``AV events''           & -0.4,+0.0& -0.4,+0.0 & $\pm0.0$ \\ \hline
Total experimental uncertainty &-7.3,+7.2 & -4.6,+3.8 &-4.3,+4.5 \\ \hline 
Cross section\cite{crosssection} & $\pm 1.1$  & $\pm 1.2 $& $\pm 0.5$ \\ \hline
\end{tabular}\label{tab:errors}}
\end{table}

\begin{figure}
\begin{center}
\centerline{\epsfxsize=3.83in\epsfbox{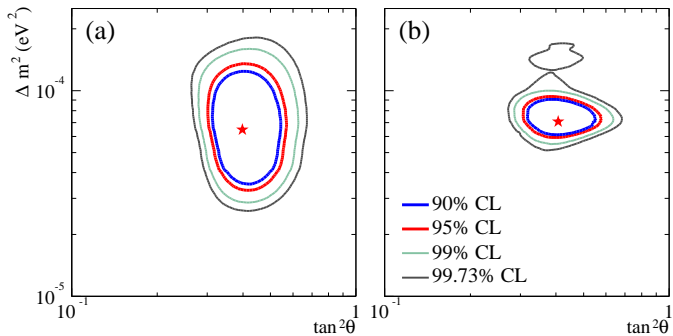}}  
\caption{\label{fig:globalmsw}Global neutrino oscillation contours. (a) Solar global: D$_2$O day and night spectra, salt CC, NC, ES fluxes, SK, Cl, Ga.  The best-fit point is $\Delta m^2=6.5\times10^{-5}$, $\tan^{2}\theta=0.40$, $f_{B}=1.04$, with $\chi^{2}$/d.o.f.=70.2/81. (b) Solar global + KamLAND.  The best-fit point is $\Delta m^2=7.1\times10^{-5}$, $\tan^{2}\theta=0.41$, $f_{B} = 1.02$.  In both (a) and (b) the ${}^{8}$B flux is free and the {\em{hep}} flux is fixed. }
\end{center}
\end{figure}

An analysis to determine the neutrino mixing parameters $\Delta m^2$ and $\tan^2 \theta$ under the assumption of two-flavour MSW neutrino oscillation was performed.  The ratio $f_{B}$ of the total ${}^8$B flux to the SSM\cite{bpb} value was a free parameter, while the total {\emph{hep}} flux was fixed at $9.3 \times 10^3$~cm$^{-2}$~s$^{-1}$.
The allowed region of the mixing parameters was determined from a combined $\chi^2$ fit to the \dto\ and salt data\cite{snocompanion}, and results from other solar neutrino experiments\cite{sage,gno,superk}.   The allowed region is shown in  Fig.~\ref{fig:globalmsw}(a).  If a further assumption of CPT invariance is made and the results from the KamLAND reactor $\bar{\nu_e}$ experiment\cite{kamland} is included, the allowed region is further reduced (Fig.~\ref{fig:globalmsw}(b)).  The best-fit point with marginalized uncertainties is $\Delta m^{2} = 7.1^{+1.0}_{-0.3}\times10^{-5}$~eV$^2$ and $\theta = 32.5^{+1.7}_{-1.6}$ degrees, which rejects the hypothesis of maximal mixing at a confidence level equivalent to \saltnsigmamax$\sigma$.

\clearpage

\section{Future Prospects and Summary}
\label{sec:3}

The SNO collaboration is now preparing for the NCD phase of its physics program.  In this phase, the NC neutrons are detected by the NCD array and the CC signals are detected by the PMTs.  This separation of the detection mechanism will significantly reduce the statistical anti-correlation of the signals.  In addition, a measurement of the neutrino energy spectrum via the CC reaction can be made with higher precision with the reduction of NC signal in the Cherenkov light spectrum.  

At the time of this talk, the \dto\ is being desalinated.  All the $^3$He counters have been constructed and are being characterized in the underground laboratory.   There are intense activities in preparing for the NCD deployment, and integration of the NCD electronics and data acquisition system.  

The SNO experiment has demonstrated conclusively in its pure \dto\ phase that solar $\nu_e$'s transform their flavour while in transit to the Earth.  A high precision measurement of the total active solar neutrino flux without any assumption of the energy dependence of the neutrino flavour transformation probability has been made in the salt phase.  In the next few months, the SNO physics program will enter its NCD phase.  It is anticipated that the data from this NCD phase, along with data from the previous phases, will significantly enhance our understanding of the properties of neutrinos.
 
\section*{Acknowledgments}
This work is supported by Canada: NSERC, Industry Canada, NRC, Northern Ontario Heritage Fund, Inco, AECL, Ontario Power Generation, HPCVL, CFI; US: DOE; UK: PPARC.  We thank the SNO technical staff for their strong contributions.

\end{document}